\documentstyle[12pt,worldsci]{article}
\begin{document}
\rightline{HEPHY-PUB 697/98}
\title{THE QUARK-ANTIQUARK WILSON LOOP FORMALISM IN THE NRQCD POWER COUNTING SCHEME}
\author{Antonio Vairo\\
{\em Institut f\"ur Hochenergiephysik, \"Osterreichische Akademie der Wissenschaften\\
 Nikolsdorfergasse 18, A-1050 Vienna, Austria}}
\maketitle
\setlength{\baselineskip}{2.6ex}
\vspace{0.7cm}
\begin{abstract}
The quark-antiquark interaction from the NRQCD Lagrangian is studied in the Wilson loop formalism. 
\end{abstract}
\vspace{0.7cm}

\section{The NRQCD Lagrangian}

Large radii quarkonia (i.e. excited heavy mesons like $\bar{b}b$ and $\bar{c}c$) 
cannot be described in terms of perturbative QCD with the addition of leading 
nonperturbative effects encoded into local condensates\cite{yndurain}. This happens  since 
in this situation the gluonic correlation length cannot be considered infinitely large with 
respect to the other scales of the system. A successful description needs therefore a systematic 
inclusion of non-local condensates. A solution is provided by the so-called Wilson loop formalism.  
The non-local quantities are field strength insertions in the Wilson loop made up by 
the quark trajectories\cite{brown,barchielli,brambilla}.  

While a full relativistic QCD formulation in this formalism is still lacking\cite{fs}, 
such a formulation is possible for heavy quark bound states which are essentially described 
by non relativistic dynamics.  These systems are characterized by a dynamical adimensional parameter, 
the quark velocity $v$, which is small and allows a classification of the energy scales 
of the problem in hard ($\sim m$), soft  ($\sim m v$) and  ultrasoft ($\sim m v^2$). 
Moreover, this provides a power counting scheme for the operators in the Lagrangian. 
The relation between $v$ and the QCD parameters is unknown (for infinitely heavy quarks 
$v$ coincides with $\alpha_{\rm s}$, for realistic quarks it is the result of 
perturbative and nonperturbative effects) but irrelevant once the power counting scheme works.  
Due to the heavy quark mass $m$, at a scale $\mu$ between $m$ 
and $m v$ the physics is still dominated by perturbative effects. Therefore, in order to describe 
heavy quark bound states, it is possible to substitute the QCD Lagrangian with an effective 
non relativistic Lagrangian via perturbative matching at that scale. The new Lagrangian is  
simpler, since the hard degrees of freedom have been integrated out explicitly, but 
equivalent to the QCD one at a given order in $\alpha_{\rm s}$ and $v$. 
This effective Lagrangian is known as the NRQCD Lagrangian\cite{lepage}. 
At order $O(v^4)$ the NRQCD Lagrangian describing a bound state between 
a quark of mass $m_1$ and an antiquark of mass $m_2$ is\cite{manohar,pineda}
\begin{eqnarray}
& & L = Q_1^\dagger\!\left(\!iD_0 + c^{(1)}_2 {{\bf D}^2\over 2 m_1} + 
c^{(1)}_4 {{\bf D}^4\over 8 m_1^3} + c^{(1)}_F g { {\bf \sigma}\cdot {\bf B} \over 2 m_1} 
+ c^{(1)}_D g { {\bf D}\!\cdot\!{\bf E} - {\bf E}\!\cdot\!{\bf D} \over 8 m_1^2} \right. 
\nonumber\\
& & \left. + i c^{(1)}_S g {{\bf \sigma} \!\!\cdot \!\!({\bf D}\!\times\!{\bf E} - {\bf E}\!\times\!{\bf D})
\over 8 m_1^2} \!\right)\!Q_1 + \hbox{\, antiquark terms}\, (1 \leftrightarrow 2)
+ {d_1\over m_1 m_2} Q_1^\dagger Q_2 Q_2^\dagger Q_1 
\nonumber\\
& & + {d_2\over m_1 m_2} Q_1^\dagger {\bf \sigma} Q_2 Q_2^\dagger {\bf \sigma} Q_1
+ {d_3\over m_1 m_2} Q_1^\dagger T^a Q_2 Q_2^\dagger T^a Q_1 
+ {d_4\over m_1 m_2} Q_1^\dagger T^a {\bf \sigma} Q_2 Q_2^\dagger T^a {\bf \sigma} Q_1, 
\label{nrqcd}
\end{eqnarray}
where $Q_j$ are the heavy quark fields.  
The coefficients $c^{(j)}_2$, $c^{(j)}_4$, ... are evaluated at the matching scale $\mu$ 
for a particle of mass $m_j$. They encode the ultraviolet regime of QCD order by order in $\alpha_{\rm s}$. 
The explicit expressions and a numerical discussion can be found in\cite{vairo}. 
The power counting rules\cite{lepage} for the operators of Eq. (\ref{nrqcd}) are $Q \sim (mv)^{3/2}$, 
${\bf D} \sim mv$, $gA_0 \sim m v^2$, $g{\bf A} \sim m v^3$, $gE\sim m^2 v^3$ and 
$gB \sim m^2 v^4$. Four quark operators which are apparently of order $v^3$ are 
actually suppressed by additional powers in $\alpha_{\rm s}$ in the matching coefficients and the 
octet contributions by an additional power in $v^2$ on singlet states.  Therefore 
in the following we will neglect these contributions with the exception of a term which 
mixes under RG transformation with the chromomagnetic operator contribution to 
the spin-spin potential\cite{chen}. We will call the corresponding matching coefficient $d$.

\section{The Wilson Loop Formalism}

The use of the Wilson loop formalism on the Lagrangian (\ref{nrqcd}) was first suggested in\cite{chen}. 
Let us sketch the derivation of the heavy quark potential. The 4-point gauge invariant Green function $G$  
associated with the Lagrangian (\ref{nrqcd}) is defined as 
$$
G(x_1,y_1,x_2,y_2) =  \langle 0 \vert  Q_2^\dagger(x_2)  \phi(x_2,x_1) Q_1(x_1)  Q_1^\dagger(y_1)  \phi(y_1,y_2) 
Q_2(y_2) \vert 0\rangle, 
$$
where $\phi(x_2,x_1) \equiv \displaystyle\exp\left\{ - ig \int_0^1 ds \, (x_2-x_1)^\mu A_\mu(x_1 + s (x_2-x_1)) \right\}$ 
is a Schwinger line added to ensure gauge invariance. After integrating out the heavy quark fields,  
$G$ can be expressed as a quantum-mechanical path integral over the quark trajectories\cite{fs}:
\begin{eqnarray*} 
& & \!\!\!\!\!\!\!
G(x_1,y_1,x_2,y_2) = \int_{y_1}^{x_1} \!\!\!{\cal D}z_1 {\cal D}p_1\int_{y_2}^{x_2} 
\!\!\!{\cal D}z_2 {\cal D}p_2 \exp \left\{i\int_{-T/2}^{T/2} \!\!dt \sum_{j=1}^2 {\bf p}_j\cdot {\bf z}_j -m_j - 
c^{(j)}_2 {p_j^2\over 2 m_j} \right. 
\\
& & \!\!\!\!\!\!\!
\quad  \left. + c^{(j)}_4 {p_j^4\over 8 m_j^3} \right\}\, 
{1 \over N_c} \left\langle {\rm Tr \, P \, T_s} \exp\left\{ -ig \oint_\Gamma dz^\mu A_\mu(z) 
+i\int_{-T/2}^{T/2} \!\!dz_{0j} \, c^{(j)}_F  g {{\bf\sigma}\cdot{\bf B}\over 2 m_j}  \right.\right. 
\\
& & \!\!\!\!\!\!\!
\quad \left.
+ c^{(j)}_D g { {\bf D}\!\cdot\!{\bf E} - {\bf E}\!\cdot\!{\bf D} \over 8 m_j^2} 
+ i c^{(j)}_S g {{\bf \sigma} \!\!\cdot \!\!({\bf D}\!\times\!{\bf E} 
- {\bf E}\!\times\!{\bf D}) \over 8 m_j^2} \right\}
\\
& & \!\!\!\!\!\!\!
\quad \left.
\times \exp\left\{ {i\over m_1m_2} \int_{-T/2}^{T/2} \!\!dt \, g^2 d\, T^{a(1)} {\bf \sigma}^{(1)} 
T^{a(2)} {\bf \sigma}^{(2)}\delta^3({\bf z}_1 - {\bf z}_2) \right\} \right\rangle
\\
& & \!\!\!\!\!\!\!
\equiv \int_{y_1}^{x_1} \!\!\!{\cal D}z_1 {\cal D}p_1\int_{y_2}^{x_2} \!\!\!{\cal D}z_2 {\cal D}p_2
\exp\left\{i\int_{-T/2}^{T/2} \!\!dt \sum_{j=1}^2 {\bf p}_j\cdot {\bf z}_j -m_j - 
{p_j^2\over 2 m_j}  + {p_j^4\over 8 m_j^3} -i \int_{-T/2}^{T/2} \!\! dt \, U\right\}, 
\end{eqnarray*}
where the bracket means the Yang--Mills average over the gauge fields, $\Gamma$ is the Wilson loop 
made up by the quark trajectories $z_1$ and $z_2$  and the endpoints Schwinger strings and 
$y_2^0 = y_1^0 \equiv -T/2$, $x_2^0 = x_1^0 \equiv T/2$. Reparameterization invariance\cite{manohar}  
fixes $c_2=c_4=1$. Assuming that the limit exists, we define the heavy 
quark-antiquark potential $V$ as $\displaystyle \lim_{T\to\infty} \int_{-T/2}^{T/2} \!\!dt \,U$. 
For a discussion on the existence of a quark-antiquark  potential we refer to\cite{lepage2,pnrqcd}.  
Expanding in $v$ (following the power counting given in the previous section) or in the inverse 
of the mass we get 
\begin{eqnarray}
& & V(r) = \lim_{T \to \infty} { i \log \langle W(\Gamma) \rangle  \over T} 
\label{pot}\\
& & \!\!\!\!\!\!\!\!\!\!\!\! + \left( {{\bf S}^{(1)}\cdot{\bf L}^{(1)}\over m_1^2} +  
{{\bf S}^{(2)}\cdot{\bf L}^{(2)}\over m_2^2} \right)\!
{2 c^+_F V_1^\prime(r) + c^+_S V_0^\prime(r) \over 2r} 
+ { {\bf S}^{(1)}\cdot{\bf L}^{(2)}  + 
{\bf S}^{(2)}\cdot{\bf L}^{(1)}  \over m_1 m_2} {c^+_F V_2^\prime(r) \over r}
\nonumber\\
& & \!\!\!\!\!\!\!\!\!\!\!\!
+ \left( {{\bf S}^{(1)}\cdot{\bf L}^{(1)}\over m_1^2} - 
{{\bf S}^{(2)}\cdot{\bf L}^{(2)}\over m_2^2} \right) \!
{2 c^-_F V_1^\prime(r) + c^-_S V_0^\prime(r) \over 2r} 
+ { {\bf S}^{(1)}\cdot{\bf L}^{(2)}  - 
{\bf S}^{(2)}\cdot{\bf L}^{(1)}  \over m_1 m_2} {c^-_F V_2^\prime(r) \over r}
\nonumber\\
& & \!\!\!\!\!\!\!\!\!\!\!\!
+{1\over 8}\left( {c_D^{(1)} \over m_1^2} 
+ {c_D^{(2)} \over m_2^2} \right) (\Delta V_0(r) + \Delta V_{\rm a}^E(r)) 
+{1\over 8}\left( {c_F^{(1)} \over m_1^2} 
+ {c_F^{(2)} \over m_2^2} \right) \Delta V_{\rm a}^B(r)
\nonumber\\
& & \!\!\!\!\!\!\!\!\!\!\!\!
+{c_F^{(1)}c_F^{(2)}\over m_1 m_2} \left( 
{{\bf S}^{(1)}\!\cdot\!{\bf r} {\bf S}^{(2)}\!\cdot\!{\bf r} \over r^2} - 
{{\bf S}^{(1)}\!\cdot \!{\bf S}^{(2)} \over 3} \right) \! V_3(r) 
+ {{\bf S}^{(1)}\!\cdot\!{\bf S}^{(2)} \over 3 m_1 m_2}
\left( c_F^{(1)} c_F^{(2)} V_4(r) -48 \pi \alpha_{\rm s} C_F \, d \, \delta^3(r)\right).
\nonumber
\end{eqnarray}
$W(\Gamma) \equiv  {\rm P \,} \displaystyle\exp\left\{ -ig \oint_\Gamma dz^\mu A_\mu(z) \right\}$ 
is the non-static Wilson loop. The expansion of it around the static Wilson loop $W(\Gamma_0)$  
($\Gamma_0$ is a $r\times T$ rectangle) gives the static potential 
$V_0 = \displaystyle\lim_{T \to \infty}  i \log \langle W(\Gamma_0) \rangle / T$ 
plus velocity (non-spin) dependent terms\cite{barchielli}. ${\bf S}^{(j)}$ and ${\bf L}^{(j)}$ 
are the spin and orbital angular momentum operators of the particle $j$. 
The matching coefficients are defined as $2 c^{\pm}_{F,S} \equiv c^{(1)}_{F,S} \pm c^{(2)}_{F,S}$. 
The ``potentials" $V_1$, $V_2$, ... are scale dependent gauge field averages of 
electric and magnetic field strength insertions in the static Wilson loop:  
\begin{eqnarray*}
\Delta V_{\rm a}^E(r) &=& 2\lim\limits_{T\to\infty}
\int_0^T\!\!dt\, \langle\!\langle {\mathbf {E}}({\mathbf 0},0){\mathbf {E}}({\mathbf 0},t)\rangle\!\rangle
- \langle\!\langle {\bf E}({\mathbf 0},0)\rangle\!\rangle ~ 
\langle\!\langle {\bf E}({\mathbf 0},t)\rangle\!\rangle, 
\\
\Delta V_{\rm a}^B(r) &=& 2\lim\limits_{T\to\infty}
\int_0^T \!\!dt\, \langle\!\langle {\mathbf {B}}({\mathbf 0},0){\mathbf {B}}({\mathbf 0},t)\rangle\!\rangle,
\\
\frac{r_k}{r}V_1'(r) &=&  \epsilon_{ijk} \!\lim\limits_{T\to\infty}\!
\int_0^T \!\!dt\,t \langle\!\langle {B}_i({\mathbf 0},0){E}_j({\mathbf0},t)\rangle\!\rangle, \\
\frac{r_k}{r}V_2'(r) &=& {1\over 2}\epsilon_{ijk}\!\lim\limits_{T\to\infty}\!
\int_0^T\!\!dt\, t\langle\!\langle {B}_i({\mathbf0},0){E}_j({\mathbf r},t)\rangle\!\rangle, \\
\left( {r_ir_j\over r^2} - {\delta_{ij}\over 3}  \right) V_3(r) &=& 2\lim\limits_{T\to\infty}
\int_0^T \!\!dt\,\left[\langle\!\langle {B}_i({\mathbf 0},0)
{B}_j({\mathbf r},t)\rangle\!\rangle
-\frac{\delta_{ij}}{3}\langle\!\langle {\mathbf {B}}({\mathbf 0},0){\mathbf
{B}}({\mathbf r},t)\rangle\!\rangle\right],\\
V_4(r) &=& 2\lim\limits_{T\to\infty} \int_0^T \!\!dt\, \langle\!\langle {\mathbf {B}}({\mathbf 0},0)
{\mathbf {B}}({\mathbf r},t)\rangle\!\rangle, 
\end{eqnarray*}
where $\langle\!\langle ~~\rangle\!\rangle\equiv\langle ~~W(\Gamma_0)\rangle /\langle W(\Gamma_0)\rangle$.

\section{Conclusions}

Two comments in order to conclude. The explicit expression for the potential (\ref{pot}) 
in terms of field strength insertions in a static Wilson loop 
is suitable for direct lattice\cite{bali} and analytic\cite{bv} evaluations. 
This makes the obtained  result of particular importance. 
Different vacuum models can be easily compared on the heavy quark potential 
predictions, once the Wilson loop average has been evaluated. 

The $O(v^2)$ leading order NRQCD Lagrangian $L = Q_1^\dagger\!\left(\!iD_0 + {\bf \partial}^2 /2 m_1 
\!\right)\!Q_1 +$ antiquark part does not contribute to (\ref{pot}) only with a static potential 
(with the exception of the perturbative contribution if evaluated in Coulomb gauge). 
Since the corresponding Wilson loop ${\rm P \,} \displaystyle\exp\left\{ -ig \oint_\Gamma dz^0 A_0(z) \right\}$ 
is a function of the non-static loop $\Gamma$, its expansion produces in general velocity 
dependent terms as well\cite{barchielli,bali,bv}. This is not surprising since the power counting 
scheme of NRQCD has to be considered as a leading order power counting scheme. 
An exact value in $v$ cannot be assigned to each term of the effective 
Lagrangian at least before the soft and ultrasoft degrees of freedom have been disentangled\cite{pnrqcd}. 

An extensive analysis of the topic discussed here can be found in\cite{vairo}, with particular 
emphasis on the relevance of the matching procedure in order to have a consistent 
potential in the perturbative regime.

\end{document}